\documentclass[aps,prb,twocolumn,superscriptaddress]{revtex4}
\usepackage{graphicx}
\usepackage{amsmath}
\usepackage{amssymb}
\usepackage{braket}
\usepackage{hyperref}
\usepackage{subfigure}
\usepackage{bm}
\usepackage[usenames,dvipsnames]{color}

\usepackage{graphicx}
\usepackage{subfigure}
\usepackage[mathscr]{euscript}

\newcommand{\ct}{\cite}

\newcommand{\bi}{\bibitem}
\newcommand{\be}{\begin{equation}}
\newcommand{\ee}{\end{equation}}
\newcommand{\ba}{\begin{eqnarray}}
\newcommand{\ea}{\end{eqnarray}}

\newcommand{\non}{\nonumber}
\newcommand{\Tr}{{\rm Tr}}

\begin{document}

\title{Bilayer Haldane system: Topological characterization and adiabatic 
passages connecting Chern phases}

\author{Sourav Bhattacharjee}
\email{bsourav@iitk.ac.in}
\affiliation{Department of Physics, Indian Institute of Technology Kanpur, 
Kanpur 208016, India}
\author{Souvik Bandyopadhyay} 
\affiliation{Department of Physics, Indian Institute of Technology Kanpur, 
Kanpur 208016, India}
\author{Diptiman Sen} 
\affiliation{Centre for High Energy Physics and Department of Physics, 
Indian Institute of Science, Bengaluru 560012, India}
\author{Amit Dutta} 
\affiliation{Department of Physics, Indian Institute of Technology Kanpur, 
Kanpur 208016, India}


\begin{abstract}
We present a complete topological characterization of a bilayer composite of two Chern insulators (specifically, Haldane models) and explicitly establish the bulk-boundary correspondences. We show that an appropriately defined Chern number accurately maps out all the possible phases of the system and remains well-defined even in the presence of degeneracies in the occupied bands. Importantly, our result paves the way for realizing adiabatic preparation of monolayer Chern insulators. This has been a major challenge till date, given the impossibility of unitarily connecting inequivalent topological phases.
We show that this difficulty can be circumvented by adiabatically varying the interlayer coupling in such a way that the system remains gapped at all times. 
In particular, a complete knowledge of the phase diagram of the bilayer composite immediately allows one to identify all such adiabatic passages which {may connect the different Chern inequivalent phases of the individual monolayers.}
\end{abstract}
\maketitle


\section{Introduction}\label{sec_intro}

The Haldane model is a paradigmatic model of two-dimensional non-interacting Chern insulators and has been subjected to extensive theoretical as well as experimental studies \ct{kane05,bernevig13,haldane83,shen12, wright13, garrity14, jotzu14, ding19}. In its commonly studied form, the model is realized on a monolayer graphene honeycomb lattice with broken sublattice and time-reversal symmetries {(see Appendix \ref{app_haldane}) for a short discussion on Haldane model)}. The topological phases of the model are characterized by an integer quantized Chern invariant; furthermore, a {topological} bulk-boundary correspondence (BBC) in the form of chiral edge states emerges in the non-trivial Chern phases. {In recent years, several works have explored composite systems of coupled Haldane layers, in particular, bilayer systems \ct{peng_cheng19,panas20,cooper20,sorn18,sen19,xiao20,huang20,shang20}}. Despite several intriguing attempts, \ct{peng_cheng19,panas20,cooper20,sorn18} it has remained unclear whether the topological structure of the monolayer Haldane system is carried over to a bilayer composite. In this regard, it has recently been shown that a `topological proximity effect' results from the gap induced in the graphene monolayer, in a coupled Haldane-graphene system \ct{peng_cheng19,panas20}. Similarly, bilayer composites of Haldane systems are known to host topological `corner states', although the edge states are gapped out \ct{sen19,xiao20,huang20,shang20}. 

In parallel, the unitary preparation or tuning of Chern insulating phases \ct{rigol15, caio15, utso17, motruk17, mardanya18, goldstein19,dutta20,dutta_floquet20} of the (monolayer) Haldane
model has remained a major challenge till date. {While there has been a fair amount of success with respect to the experimental preparation of materials hosting Chern non-trivial phases\ct{wright13, garrity14, jotzu14, ding19}, dynamical tuning or switching across the different Chern phases in a given Chern insulator is altogether a different challenge.
{To elaborate, the difficulty is two-fold. First, one needs to engineer the effective Hamiltonian generating the time evolution of the system in such a 
way that the ground state of the engineered Hamiltonian is in the desired topological phase. This, for example, can be achieved simply through a sudden quench or a periodic modulation of the Hamiltonian. In this regard, it has been demonstrated that the effective Floquet Hamiltonian driving the stroboscopic dynamics of a periodically modulated system can host topologically non-trivial phases, despite the ground state of the undriven Hamiltonian being in a trivial/non-topological phase.} This idea has also been exemplified through the application of circularly polarized radiation on graphene (see Refs.~\onlinecite{oka09,kitagawa_demler11}).} 
	
{{Secondly}, {in generic out-of-equilibrium systems, the time-evolved 
many-body state is not an eigenstate of the effective Hamiltonian generating the dynamics. Thus, the time-evolved state may not exhibit a topological BBC, as expected from the non-trivial effective Hamiltonian}. An immediate and apparent 
solution to the problem is to maintain adiabaticity throughout the 
dynamics, at least in incommensurate finite-size systems,
i.e., systems in which the gapless point is excluded from the Brillouin zone
(see Ref.~\onlinecite{ge21}). This ensures that the out-of-equilibrium state closely follows the ground state of the effective Hamiltonian. It has indeed been shown that by maintaining adiabaticity in finite-size systems, the lattice Chern number or the Bott topological index \cite{rigol15} can capture a dynamical topological phase transition in the non-equilibrium state of the system. This is, nevertheless, a difficult task to achieve experimentally as the dynamics needs to be extremely slow for sufficiently large systems and therefore requires a long coherence time of the system.}
	
{It is {also} important to realize that a topological bulk-boundary 
correspondence (BBC) only holds in the thermodynamic limit, i.e., when the conducting edge states decaying exponentially\ct{hasan10} into the bulk, do not hybridize. }{One therefore must address the dynamical preparation or tuning of non-trivial Chern states not just in finite systems but also in the thermodynamic limit. However, maintaining adiabaticity in dynamics to tune the system across different Chern phases is impractical for {thermodynamically} large systems, as the minimum energy gap vanishes for such systems at the critical points separating inequivalent topological sectors. In fact, it has already been established \ct{ge21} that, neither the Chern number nor the Bott invariant can be changed under unitary dynamics in thermodynamically large or commensurate (i.e., systems in which the gapless point is included within the Brillouin zone) translationally invariant monolayer Chern insulators (see Ref.~\onlinecite{sen21} for a deeper discussion on the problem).}\\


{{In this work, we address the aforementioned obstacle in the dynamical preparation and tuning of non-trivial Chern states in the thermodynamic limit. To this end, we consider the possibility} of adiabatically tuning the phase of a thermodynamically large Chern insulator when coupled to a similar, but not necessarily identical, Chern insulating system. To elaborate, we first analyze the topological characterization of a bilayer Haldane system in its ground state and the associated BBCs.} We find that even in the presence of a finite coupling between the layers, the bilayer system is capable of hosting topologically non-trivial phases, although the topology of the individual layers cease to be well-defined. In particular, we make use of the total Chern number, calculated from the non-Abelian Berry curvature \ct{fukui05, wilczek84, murakami04, nayak08,goldman19,belzig20} to identify the topological phases of the bilayer system. As our main result, we show that such bilayer composites facilitate unitary preparation or tuning of Chern phases {in commensurate and incommensurate monolayer Chern insulating lattices.} \\

\begin{figure}[t]
	\subfigure[]{
		\includegraphics[width=0.5\columnwidth]{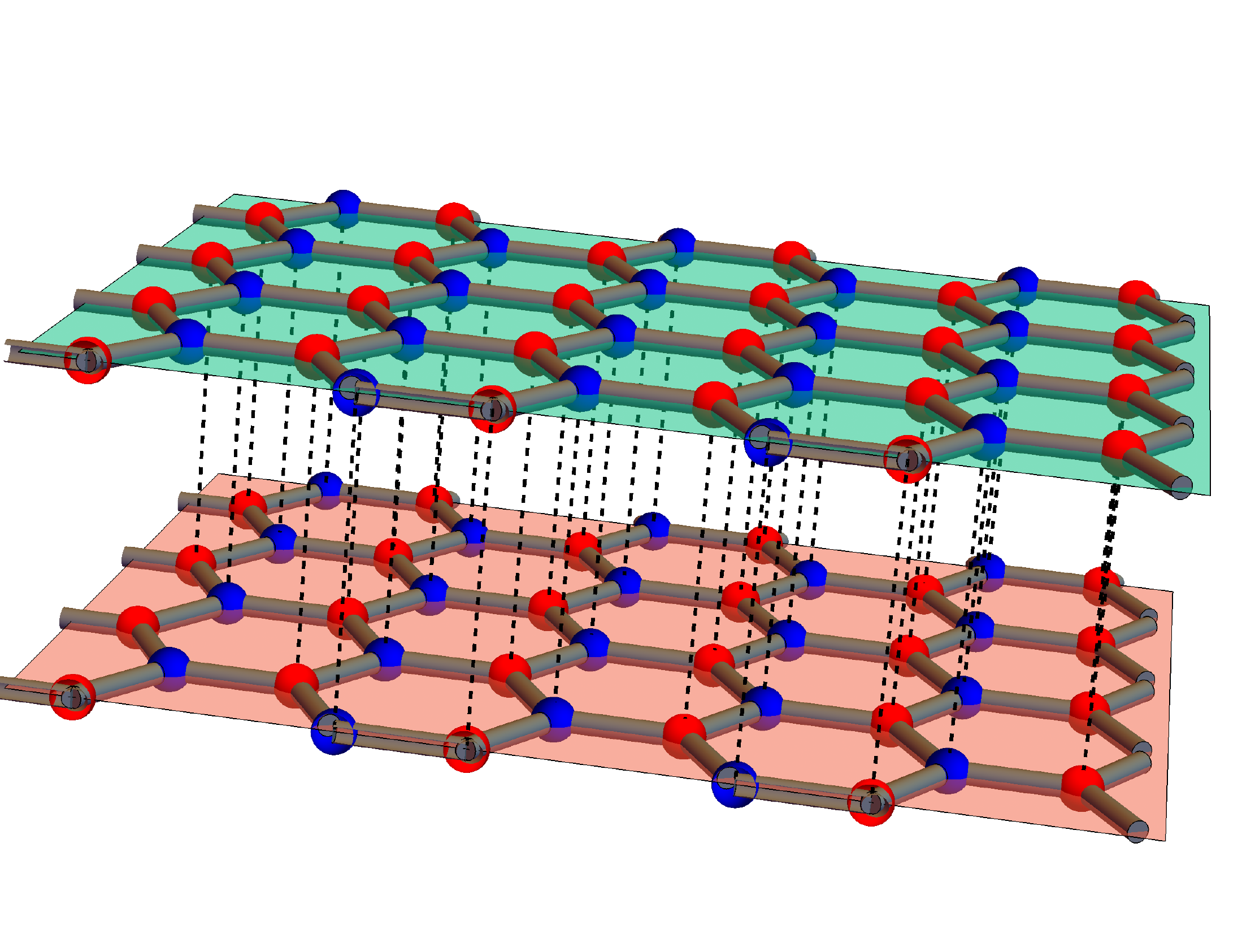}
		\label{fig_bilayer}}%
	\subfigure[]{
		\includegraphics[width=0.5\columnwidth]{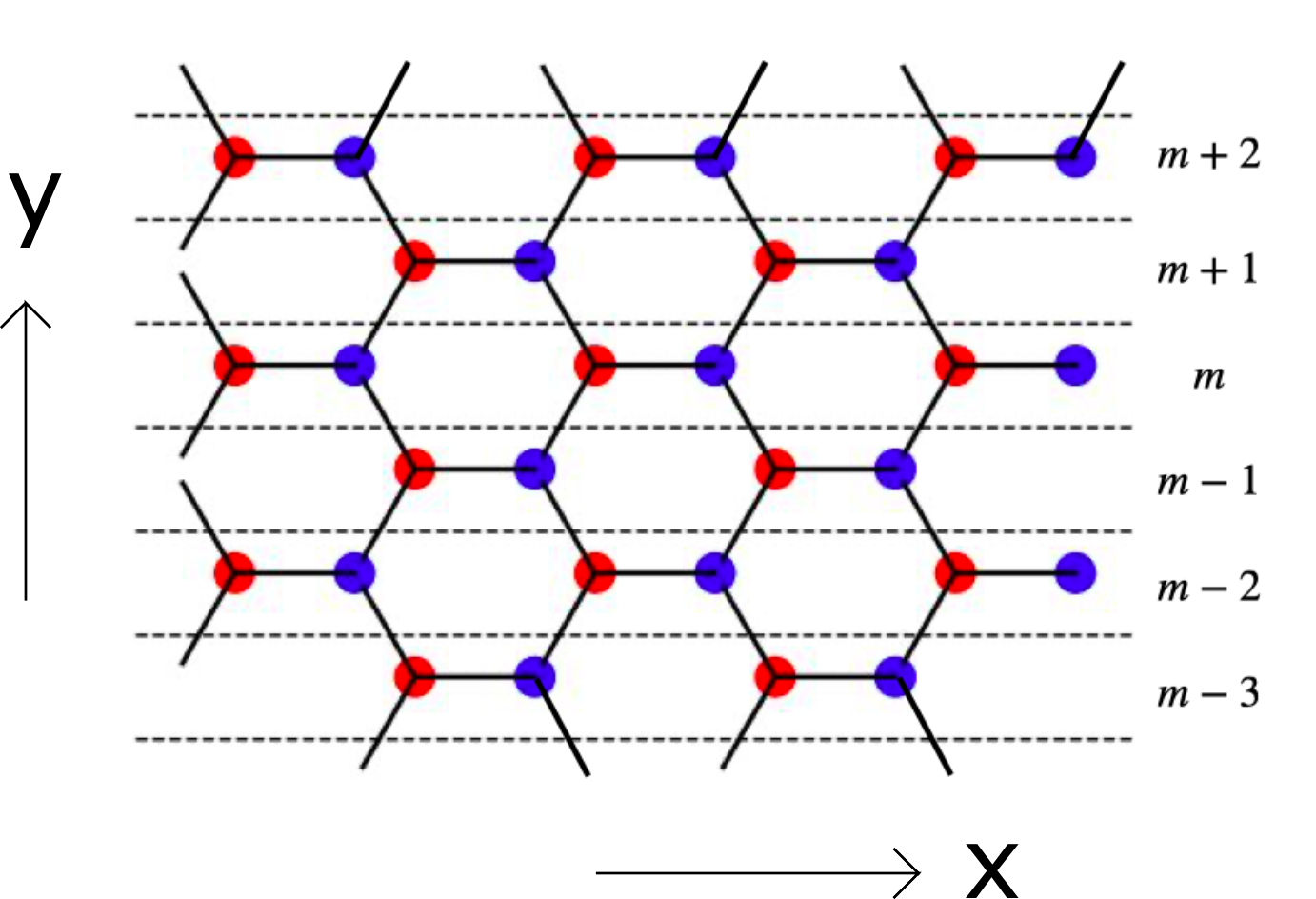}
		\label{fig_strip}}	
	\caption{(a) The bilayer Haldane model realized by two vertically stacked and perfectly aligned honeycomb lattices. The red and blue spheres correspond to the two sublattices and the black dashed lines indicate that each lattice point is coupled only with the one directly above or below it. (b) (Top view) -- In a semi-infinite system with armchair edges, the lower (upper) layer can be divided into $M$ `strips' of chains, each indexed by the letter $m_{l(u)} = 1,~2,~3,\dots, M$.}\label{fig_setup}
\end{figure}

To this end, a complete knowledge of the topological phases of the bilayer composite is crucial to identify \textit{adiabatic passages}, which can be traversed to tune the Chern phases (defined in the absence of interlayer coupling) of the individual layers. {The adiabaticity ensures that the bulk-boundary correspondence is restored at the end of the tuning process.} Importantly, the {adiabatic passages} persist even in the thermodynamic limit, a finding which is highly significant in the context of {unitarily tuning the phases of Chern insulators. As we discuss in the conclusion, the dynamical coupling to such an ancillary layer also allows one to search for optimal adiabatic pathways aiding the experimental realization of such protocols.}


\begin{figure*}%
	\subfigure[]{
	\includegraphics[width=0.2\textwidth]{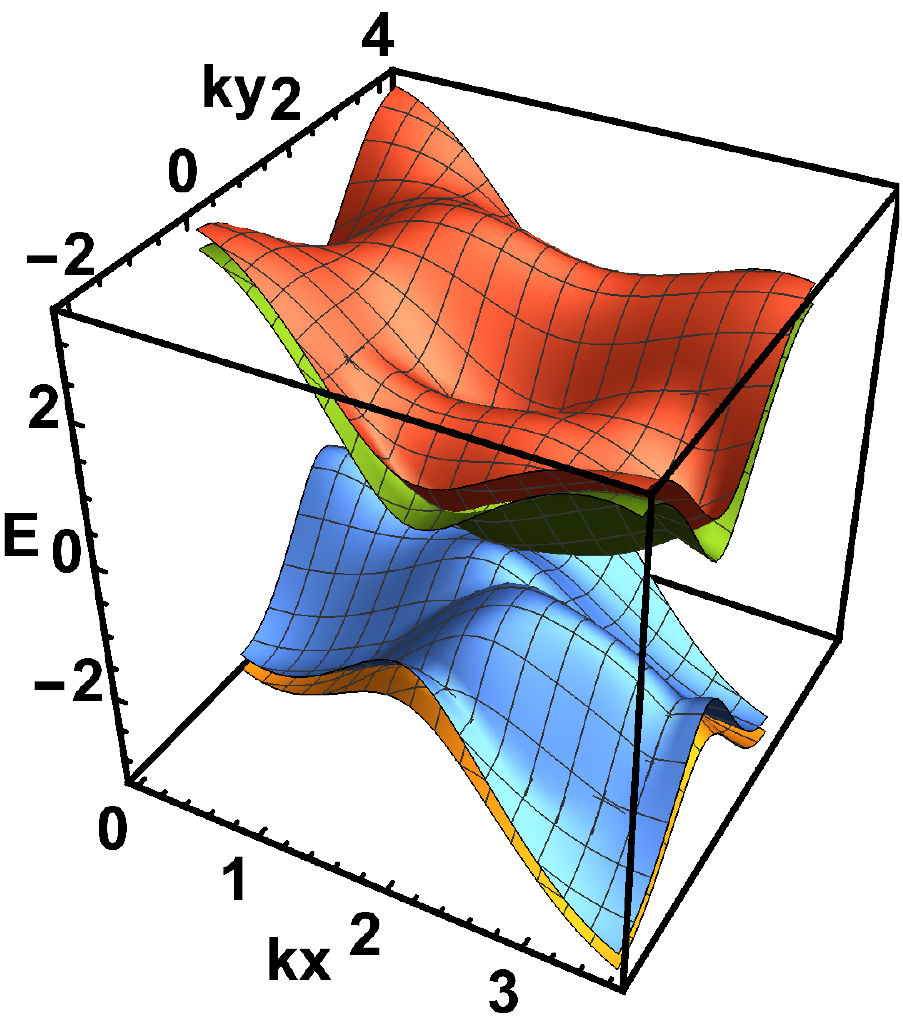}
	\label{fig_disp}}%
	\subfigure[]{
	\includegraphics[width=0.3\textwidth]{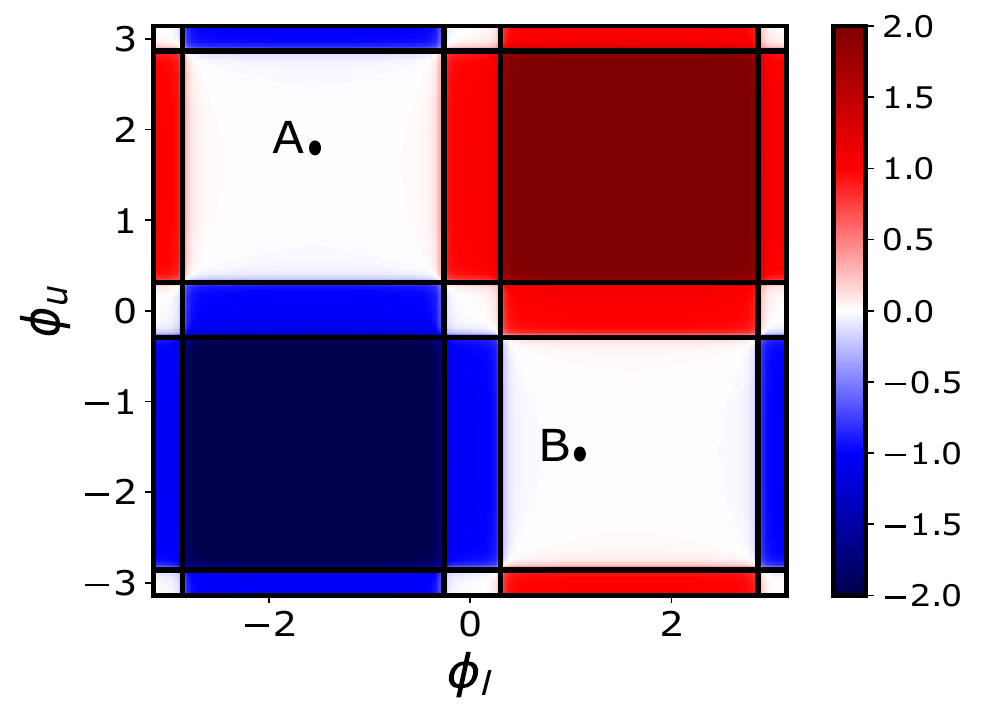}
	\label{fig_gamma0}}
	\subfigure[]{
	\includegraphics[width=0.3\textwidth]{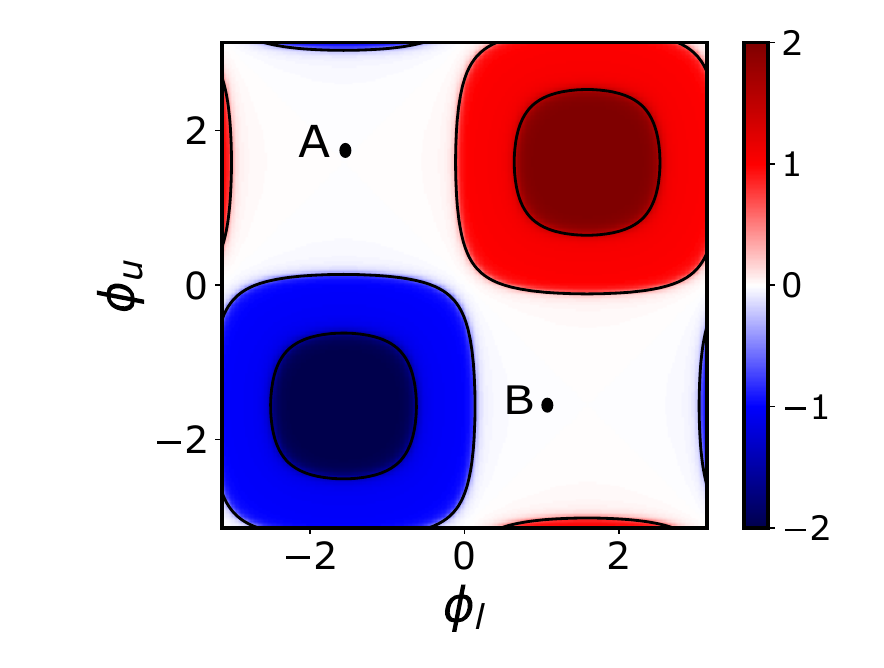}
	\label{fig_phase}}%
	\caption{(a)~The four dispersion bands of the bilayer Haldane model corresponding to the Hamiltonian in Eq.~\eqref{eq_hamil}. Topological phases of the bilayer Haldane model in the $\phi_l-\phi_u$ plane {with (b) $\gamma = 0$ and (c) $\gamma = 0.8$}. The black solid lines represent the critical boundaries between the different phases at which the bulk gap vanishes. The colors indicate the values of the total Chern number which acquire only integer quantized values, ranging from $-2$ to $2$. {Note that the points A and B are no longer separated by critical lines when $\gamma\neq 0$}. The other parameter values chosen for the plots are $t_1=1$, $t_2=1/3$ and $M=0.5$.}
\end{figure*}

\section{Model}\label{sec_model}

In our model we make the simplifying assumption that the two layers have identical sets of values of the Semenoff mass $M$ as well as the nearest-neighbor (NN) and the next-nearest-neighbor (NNN) hopping amplitudes, $t_1$ and $t_2$, respectively. However, they may differ with respect to the phase of the complex NNN hoppings. We will denote the corresponding phases of 
the `lower' and the `upper' layers as $\phi_l$ and $\phi_u$, respectively. In addition, the interlayer interaction is so chosen, that within the translationally invariant bulk, modes with different lattice momenta $\bm k$ in the Brillouin zone (BZ) do not couple. This retains the integrability of the composite system.\\

Assuming periodic boundary conditions for the bulk, the Hamiltonian is decoupled as, $H=\bigoplus_{\bf k} {\bm c}_{\bf k}^\dagger H(\bm k){\bm c}^{}_{\bf k}$, where ${\bm c}_{\bf k} = \left(c_{{\bf k},\mathrm{A}}^l, c_{{\bf k},\mathrm{B}}^l, c_{{\bf k},\mathrm{A}}^u, c_{{\bf k},\mathrm{B}}^u\right)$ is a vector of the annihilation operators with $\{A,B\}$ and $\{l,u\}$ being the sublattice and layer indices, respectively. The single-particle Hamiltonian $H(\bm k)$ assumes the form, 
\begin{equation}\label{eq_hamil}
H(\bm k)=\bigoplus_{\bf k}\begin{pmatrix}
		H_l(\bm k) & \Gamma\\
		\Gamma^\dagger & H_u(\bm k)
			\end{pmatrix},
\end{equation}
where $H_{l(u)}$ is the Haldane Hamiltonian corresponding to the lower (upper) layer and $\Gamma$ denotes the interaction potential between the layers. We recall the Bloch form of the Haldane Hamiltonians {(see Appendix \ref{app_haldane})}, $H_{l(u)}(\bm k)={\bm d}^{l(u)}(\bm k)\cdot{\bm \sigma}$, where $\bm d^{l(u)}=\{d_x,d_y,d_z^{l(u)}\}$ and $\bm \sigma$ is a vector of pseudo-spin operators. Note that only $ d_z^{l(u)}$ depends on the complex phase and is therefore annotated with distinct superscripts for each layer. In what follows, we consider a staggered interlayer coupling of the form $\Gamma=\gamma\tau_z$, where $\tau_z$ is another pseudo-spin operator. Physically, such a situation may arise when the two graphene sheets are perfectly aligned with each other and satisfying the following two conditions: (i) each lattice point in the upper layer interacts only with the lattice point directly below it in the lower layer (see Fig.~\ref{fig_bilayer}), and (ii) the interaction is attractive or repulsive depending on which of the two sublattices a given point belongs to. We emphasize here that the staggered nature of the interaction (condition (ii)) only simplifies the analysis of the topological phases and our results remain qualitatively unaltered for more general interactions {as discussed in Sec.~\ref{sec_phase}.}

 
Analyzing the spectrum of the Hamiltonian in Eq.~\eqref{eq_hamil}, the energy bands assume the form,
\begin{widetext}
\begin{subequations}\label{eq_spec}
	\begin{equation}
	E_1^\pm(\bm k)=\pm\sqrt{d_x^2(\bm k)+d_y^2(\bm k)+\frac{1}{4}\left(d_z^l(\bm k)+d_z^u(\bm k)+\sqrt{(d_z^l(\bm k)-d_z^u(\bm k))^2+4\gamma^2}\right)^2},
	\end{equation}
	\begin{equation}
	E_2^\pm(\bm k)=\pm\sqrt{d_x^2(\bm k)+d_y^2(\bm k)+\frac{1}{4}\left(d_z^l(\bm k)+d_z^u(\bm k)-\sqrt{(d_z^l(\bm k)-d_z^u(\bm k))^2+4\gamma^2}\right)^2},
	\end{equation}
\end{subequations} 
\end{widetext}
where $E_1^-(\bm k)\leq E_2^-(\bm k)\leq 0 \leq E_2^+(\bm k)\leq E_1^+(\bm k)$ (see Fig.~\ref{fig_disp}). In the ground state, only $E_1^-$ and $E_2^-$ are completely occupied while the rest are completely empty. Clearly, a finite gap between the occupied and empty band ensures that the bulk of the system remains insulating. Note that the efficacy of choosing a staggered interaction, as manifested in Eqs.~\eqref{eq_spec}, is that the bulk gap can vanish only at the Dirac points $\bm K,~\bm K'$, where $d_x(\bm K,\bm K')=d_y(\bm K,\bm K')=0$. Hence it suffices to analyze the 
spectrum in the vicinity of the Dirac points only. In particular, the critical points are found by setting $E_2^{\pm}(\bm K)=0$ and $E_2^{\pm}(\bm K')=0$, leading to the conditions,
	\begin{align}\label{eq_critic_bound}
	d_z^l(\bm K)d_z^u(\bm K)=\gamma^2,\quad\quad
	d_z^l(\bm K')d_z^u(\bm K')=\gamma^2.
	\end{align}

In the limiting case $\gamma=0$, the conditions in Eq.~\eqref{eq_critic_bound} are satisfied when $d_z^{l(u)}(\bm K)=0$ and/or $d_z^{l(u)}(\bm K')=0$, implying that at least one of the independent Haldane layers undergoes a topological phase transition. This is trivially expected since the 
topological properties of the composite system can be deduced from that of the individual layers, in terms of the Chern numbers $C_{l(u)} =0, \pm1$ of the lower (upper) layers. \\

The situation is however not trivial for $\gamma\neq 0$ since the finite 
interaction between the layers no longer guarantees particle number 
conservation of the individual layers. In Fig.~\ref{fig_phase}, the critical boundaries (black solid lines) obtained from Eq.~\eqref{eq_critic_bound} are plotted in the $\phi_l - \phi_u$ plane for fixed values of $M$, $t_1$ and $t_2$. The critical lines separate the $\phi_l-\phi_u$ plane into distinct regions which, as we will demonstrate below, are characterized by integer quantized topological invariants.

\section{Phase diagram and BBC}\label{sec_phase}

For systems with multiple occupied bands, the Chern invariant characterizing the topological phases is calculated from the $U(2)$ Berry curvature which is non-Abelian {(see Appendix \ref{app_chern} for a detailed discussion)}. The Chern number thus defined turns out to be equivalent to the total Chern number calculated by summing up the Abelian curvatures of the individual bands. It is given by,
\begin{equation}\label{eq_chern}
C_{tot} = \frac{i}{2\pi}\int_{BZ}d^2k~\Tr\left(P\left[\frac{\partial P}{\partial k_i}, \frac{\partial P}{\partial k_j} \right]\right), 
\end{equation}
Here, $P = \sum_{n=1}^{2}\ket{n(\bm k)}\bra{n(\bm k)}$ is the projection operator on the ground state manifold of occupied states, with $\ket{n(\bm k)}$ being the occupied energy eigenstates of $H(\bm k)$. It is important to note that the total Chern number, as defined in Eq.~\eqref{eq_chern}, remains integer quantized and well-defined as long as the gap between the occupied and the empty bands remains finite. This includes situations in which the occupied bands may become degenerate at some points on the BZ.\\

\begin{figure*}[t]
	\centering
	\subfigure[]{
		\includegraphics[width=0.33\textwidth, height=5cm]{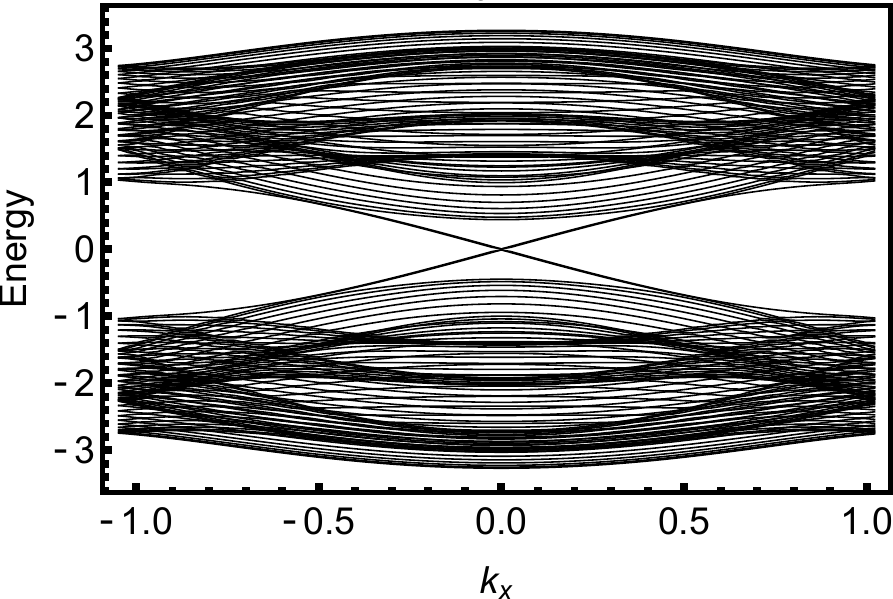}
		\label{fig_c2}}	
	\subfigure[]{
		\includegraphics[width=0.3\textwidth, height=5cm]{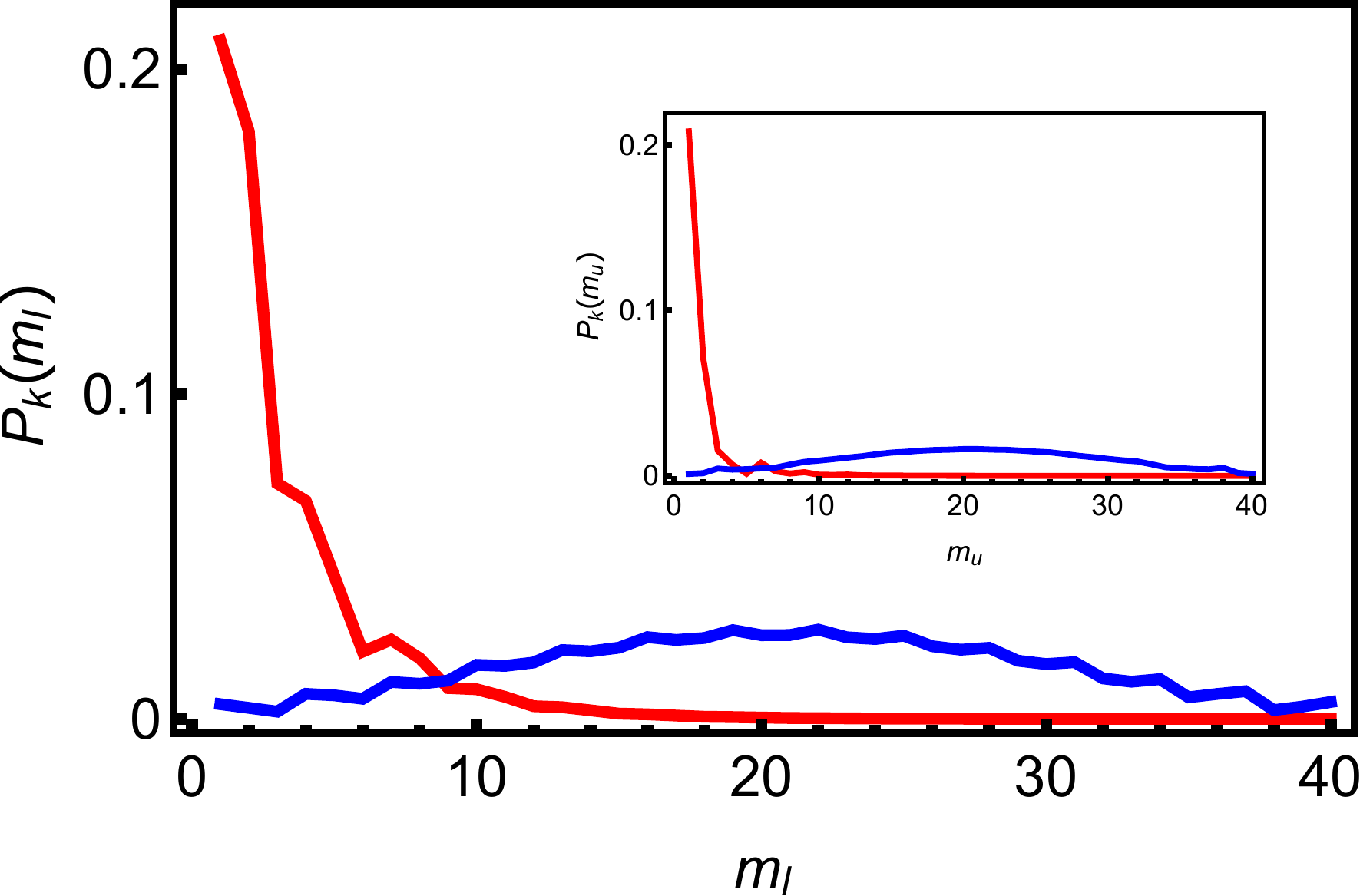}
		\label{fig_c1_u}}
	\subfigure[]{
		\includegraphics[width=0.3\textwidth, height=5cm]{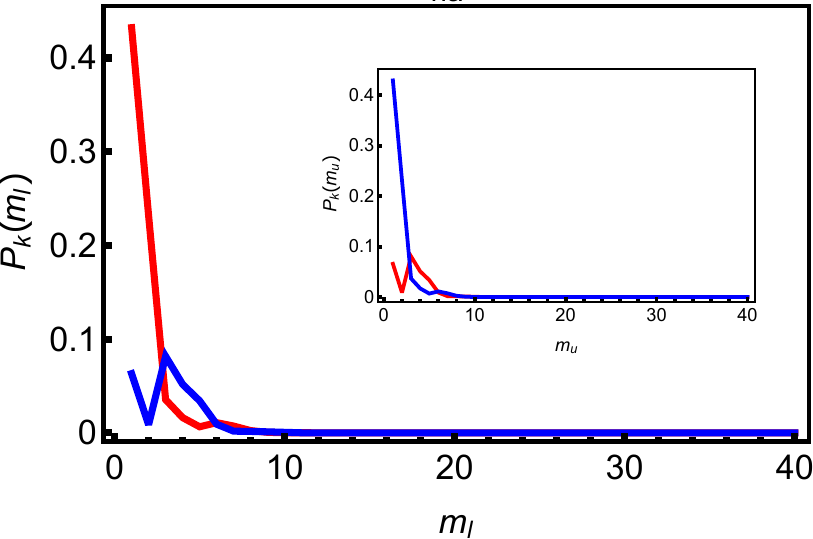}
		\label{fig_c2_u}}	
	
	\caption{(a) Spectrum of the semi-infinite bilayer Haldane model with armchair edges and $C_{tot}=2$. The values of the complex phases are chosen to be $\phi_l=1.55,~\phi_u=1.59$. The other parameters are 
		the same as in Fig.~\ref{fig_phase}. The probability distribution $P_k(m_{l(u)})$ (see Eq.~\eqref{eq_prob}) for two of the highest energy modes in the valence band as a function of $m_{l(u)}$ in the cases of (b) $C_{tot}=1$ and (c) $C_{tot}=2$. In the former case, only one localized edge mode, shown in red, exists which spans the edge of both the lower and the upper (inset of (b)) layers, while the other mode, shown in blue, diffuses into the bulk. On the other hand, for $C_{tot}=2$, two edge modes exist with one of them (red) localized at the edge of the lower layer and the other (blue) at the edge of the upper layer (inset of (c)).}\label{fig_spectrum}
\end{figure*}

Fig.~\ref{fig_phase} illustrates that the regions separated by the critical lines acquire distinct values of the Chern number, $C_{tot} = 0, \pm1, \pm2$, suggesting that the bilayer Haldane system is endowed with a rich topological structure even in the presence of a finite interaction between the layers. The natural question which then arises is whether there exists any BBC corresponding to the integer quantized values of $C_{tot}$. To this end, 
we consider the case of a semi-infinite bilayer composite, where the system is infinite along the cartesian $x$-axis and has a finite width along the $y$-axis. Exploiting the conservation of $k_x$, we depict the resulting spectrum for different values of Chern number in Fig.~\ref{fig_spectrum}. It is evident that when $C_{tot}=0$, the energy spectrum is gapped and no conducting edge states exist. On the contrary, when $C_{tot}\neq 0$, conducting edge modes appear in the bulk gap, connecting the filled valence band with the empty conduction band. 

To further establish the BBC, we inspect the probability distribution of the ground state of the Hamiltonian 
corresponding to a particular lattice momentum $k_x$, along the finite $y$-axis of each of the layers. As shown in Fig.~\ref{fig_strip}, we can divide each of the layers into $M$ `strips' along $y$ for armchair boundary edges. For a given energy eigenstate $\ket{\psi(k_x=k)}$, we then calculate the quantity given by
\begin{equation}\label{eq_prob}
P_{k}(m_{l(u)}) = \sum_{s=1}^2\left|\braket{m_{l(u)}^s|\psi(k)}\right|^2,
\end{equation}
where $m_{l(u)}^s$ is a lattice point on the $m^{th}$ strip of the lower (upper) layer, and $s$ labels the sublattice to which the lattice point belongs.

In Figs.~\ref{fig_c1_u} and~\ref{fig_c2_u} , we plot the quantity defined in Eq.~\eqref{eq_prob} as a function of $m_{l(u)}$ for two of the highest energy occupied eigenstates at a
lattice momentum $k_x>0$. For $C_{tot}=1$ (see Fig.~\ref{fig_c1_u}), we see that only one of the eigenstates is localized at the edges and it spans the edges of both the layers. On the other hand, for $C_{tot}=2$ (see Fig.~\ref{fig_c2_u}), it is evident that both eigenstates are edge-localized with each of them spanning the edge of only one of the two layers. Thus, we see a direct correspondence between the value of $C_{tot}$ and the
number of localized edge states. This correspondence is also corroborated by calculating the inverse participation ratios (IPR) of the energy eigenvalues {as shown in Appendix~\ref{app_ipr}}. Further, we have verified that the edge states localize at the opposite edge of the layers for $k_x <0$, thus establishing 
their chiral nature. Interestingly, we note that the Chern number turns out to be identical to the Chern number of the lowest energy band of two-particles energy eigenstates {see Appendix~\ref{app_2p} for details}.

\section{Unitarily connecting inequivalent phases of monolayer Chern insulators}

As already mentioned, the presence of critical boundaries separating the topological phases of a monolayer Haldane system makes it impossible to tune the system across different phases. However, we have already seen that the presence of finite inter-coupling alters the phase space structure {(see Figs.~\ref{fig_gamma0} and~\ref{fig_phase})}, which can open up adiabatic passages connecting distinct topological phases (defined for $\gamma=0$), of the monolayers. To exemplify this, we will demonstrate a simple case in which an initially decoupled bilayer system with $C_l^i=-C_u^i=1$ is adiabatically transformed to another decoupled configuration, $C_l^f=-C_u^f=-1$. Thus the protocol exchanges the Chern numbers of the layers at the end of the process. {The underlying idea is to dynamically break the ${\rm U(1)}\times{\rm U(1)}$ sub-group of the complete ${\rm U(2)}$ gauge symmetry so as to facilitate the adiabatic tuning of the monolayer Chern phases, followed by complete restoration of the same symmetry.} The transformation is achieved through an appropriate manipulation of the parameters $\gamma$, $\phi_l$ and $\phi_u$, such that adiabatic conditions are maintained throughout the process. \\

We now outline the protocol which is a three-step process. {We assume that the two layers are initialized in the ground state with $\gamma = 0$ so that each of them has well-defined Chern numbers. The other Hamiltonian parameters are so chosen such that the Chern numbers for the lower and the upper layers are $C_l^i=1$ and $C_u^i=-1$, respectively.} In the first step of the protocol, the interlayer coupling $\gamma$ is slowly ramped to a finite value, so as to open up adiabatic passages between the desired initial and final configurations. {This is exemplified in Figs.~\ref{fig_gamma0} and~\ref{fig_phase}, where the points A and B are no longer separated by critical lines in the presence of a finite $\gamma$}. In the next step, the complex phases $\phi_l$ and $\phi_u$ are slowly tuned to their target values, keeping $\gamma$ constant. {Note that in the absence of the inter-layer coupling, this would have required crossing the critical lines which cannot be performed adiabatically in the thermodynamic limit. In the final step, the interlayer coupling is slowly turned off so that the two layers once again acquire well-defined Chern numbers at the end of the protocol. Since adiabaticity is maintained throughout, the two layers are expected to remain in their respective ground states.}
The three-step protocol is encoded in the time-dependence of the Hamiltonian parameters as,
\begin{subequations} \label{eq_proto}
\begin{equation}\label{eq_proto_gamma}
\gamma(t) = \left\{
\begin{array}{ll}
\gamma_c\frac{t}{\tau'} & \mbox{for }0\leq t \leq \tau' \\
\gamma_c & \mbox{for } \tau' \leq t \leq \tau'' \\
\gamma_c\left(1-\frac{t-\tau''}{\tau-\tau''}\right) & \mbox{for }\tau''\leq t \leq \tau
\end{array}
\right.
\end{equation}
\begin{equation}
\phi_{l(u)}(t) = \left\{
\begin{array}{ll}
\phi_{l(u)}^i & \mbox{for } t \leq \tau' \\
\phi_{l(u)}^i + (\phi_{l(u)}^f-\phi_{l(u)}^i)\frac{t-\tau'}{\tau''-\tau'} & \mbox{for } \tau' \leq t \leq \tau'' \\
\phi_{l(u)}^f & \mbox{for } t \geq \tau''
\end{array}
\right.
\end{equation}
\end{subequations}

\begin{figure}[t]%
	\includegraphics[width=0.8\columnwidth]{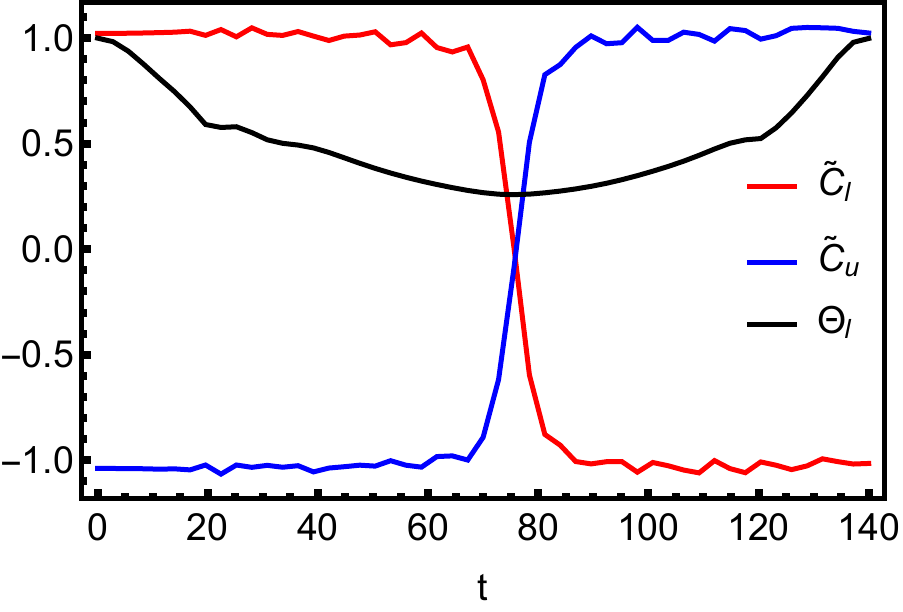}
	\caption{Dynamical exchange of the Chern numbers between the two layers attained using the protocol in Eq.~\eqref{eq_proto}. The Chern numbers (red and blue curves) are determined using Eq.~\eqref{eq_dyn_chern} and do not have any BBC in the intermediate stage, $0<t<\tau$. At the end of the protocol, the purity of the states representing the layers, shown only for the lower layer above (black curve), are restored to unity with $\tilde{C}_{l(u)}(\tau)=C_{l(u)}$. Note that the total Chern number remains invariant throughout the process $C_{tot}=0$. The parameter values chosen are $t_1 =1, t_2=1/3, M=0.5, \phi_{l}^i=-\phi_{l}^f = 0.5, \phi_u^i=-0.9, \phi_u^f=0.6, \gamma_c = 1, \tau'=20, \tau''=120$ and $\tau=140$.}\label{fig_dyn}
\end{figure}

\begin{figure*}
	\subfigure[]{
		\includegraphics[width=0.4\textwidth]{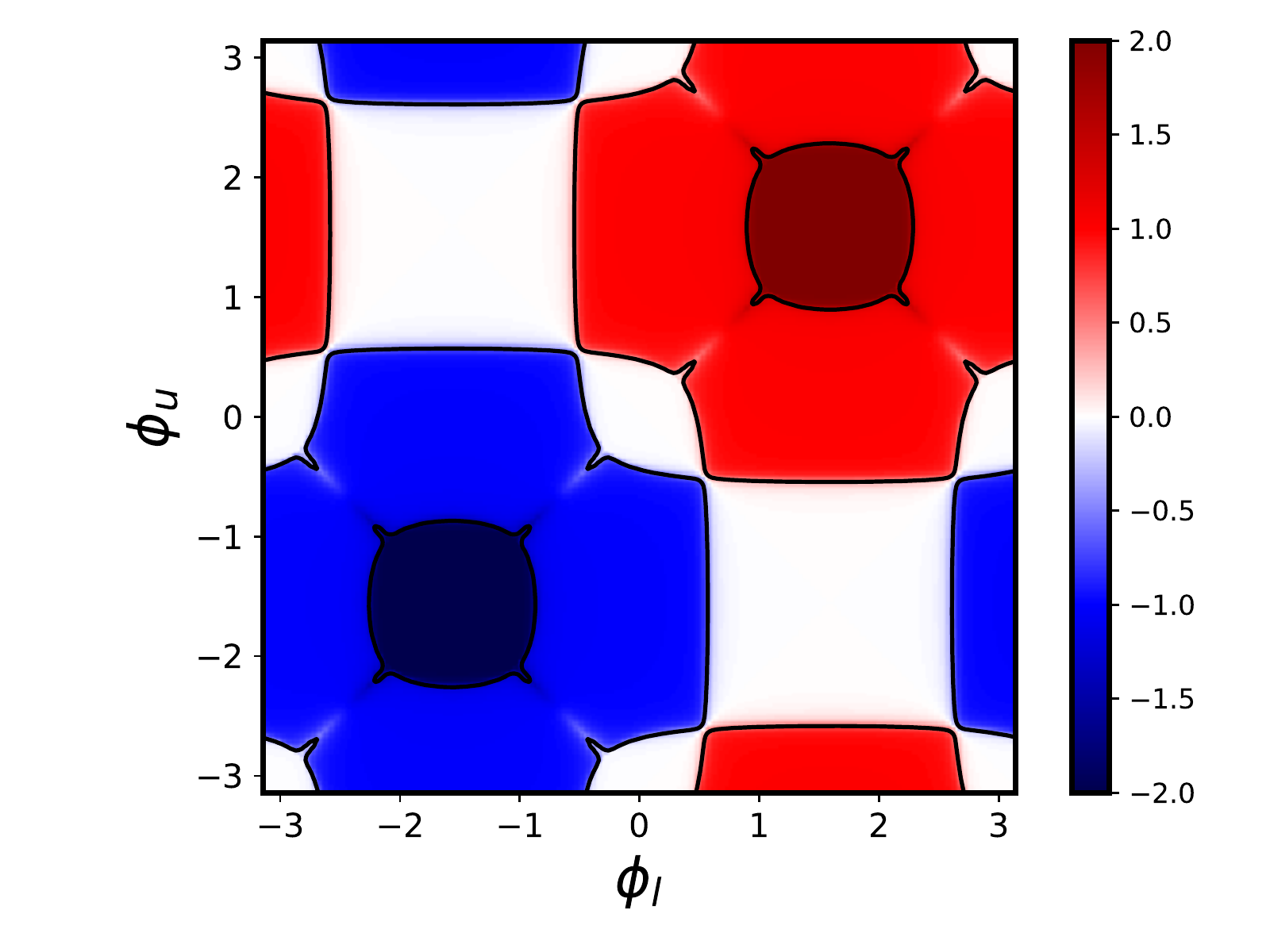}
		\label{fig_norm_phase}}\quad%
	\subfigure[]{
		\includegraphics[width=0.4\textwidth]{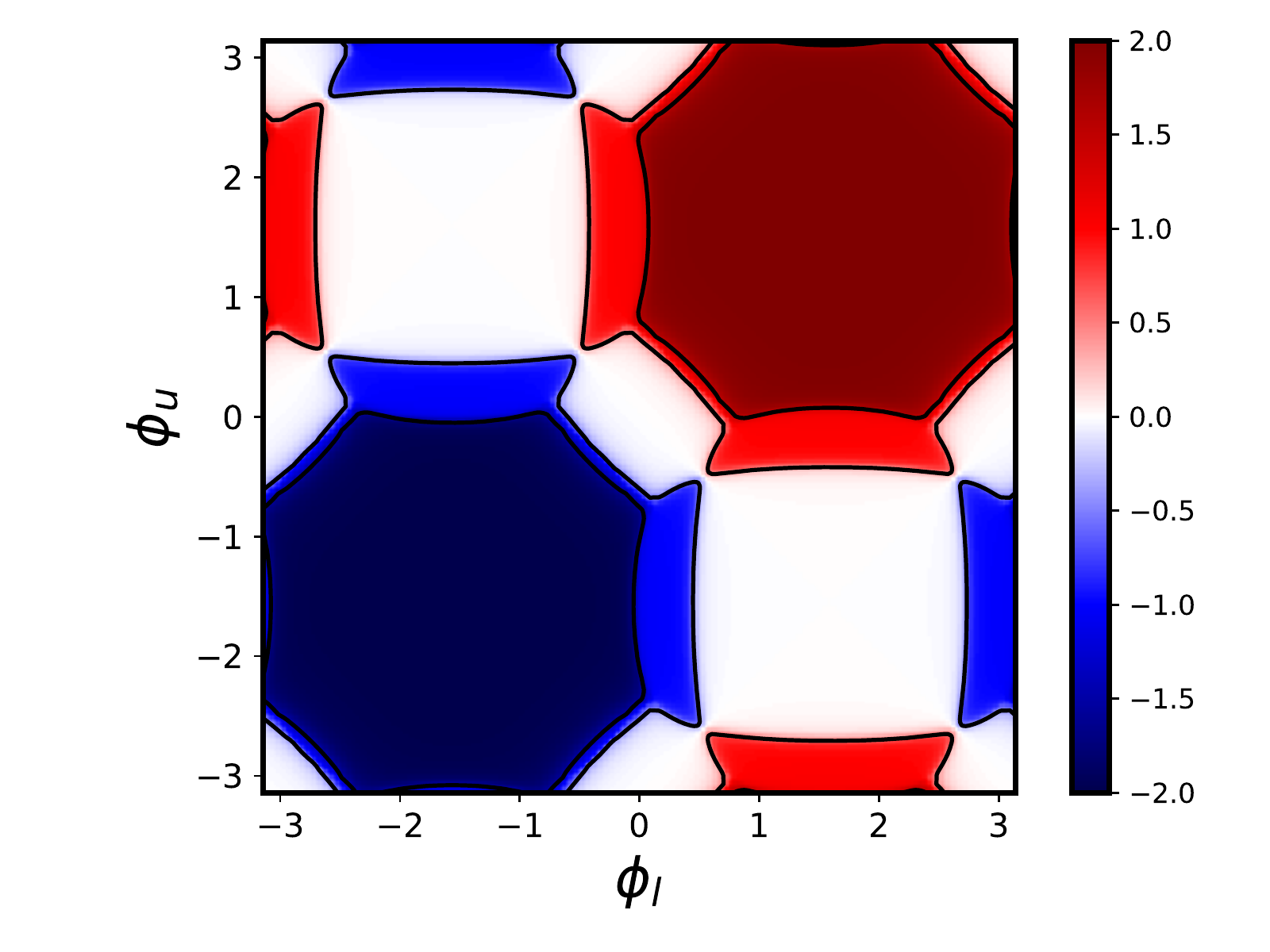}
		\label{fig_inv_phase}}	
	\caption{Phase diagram in the $\phi_l-\phi_u$ plane for (a) $\Gamma=\gamma\mathbb{I}$ and (b) $\Gamma=\gamma\tau_x$. In all cases, one can identify distinct topological phases with $C_{tot}=0,\pm1,\pm2$.}\label{fig_gen_phase}. 
\end{figure*}
At intermediate times, the presence of a finite coupling implies that the layers are entangled and hence they do not have well-defined Chern numbers. Nevertheless, it is possible to define a Chern number (see for example Ref.~\onlinecite{budich15}) in terms of the reduced state of the individual layers, $\rho_{l(u)}(\bm k)=\Tr_{u(l)}\left(\ket{\psi(\bm k)}\bra{\psi(\bm k)}\right)$, where $\ket{\psi(\bm k)}$ represents the time-evolved state of the bilayer system for the momentum $\bm k$ (the momentum modes always remain decoupled). Defining the operators, $\Lambda_{l(u)} = \Xi\rho_{l(u)}/\mathcal{N}$, where the operator $\Xi$ projects the density matrix into the single-particle subspace and $\mathcal{N}=\sqrt{2\Tr\left[\rho_{l(u)}^2\right]-\Tr\left[\rho_{l(u)}\right]^2}$is a normalization constant, a time-dependent Chern number of each layer can then be defined as,
\begin{equation}\label{eq_dyn_chern}
\tilde{C}_{l(u)}(t) = \frac{i}{2\pi}\int_{BZ}d^2k~\Tr\left(\Lambda_{l(u)}(t)\left[\frac{\partial \Lambda_{l(u)}(t)}{\partial k_i}, \frac{\partial \Lambda_{l(u)}(t)}{\partial k_j} \right]\right). 
\end{equation}
It is straightforward to check that the time-dependent Chern number defined above reduces to the exact Chern number of the individual layers defined in Eq.~\eqref{eq_chern} when $\rho_{l(u)}=\ket{g_{l(u)}(\bm k)}\bra{g_{l(u)}(\bm k)}$, where $\ket{g_{l(u)}(\bm k)}$ is the ground state of the lower (upper) layer corresponding the momentum mode $\bm k$. We assume this to be true at $t=0$, such that $\tilde{C}_{l(u)}(0)=C_{l(u)}^i$. \\

For our demonstration, we assume the following parameter values: $t_1 =1, t_2=1/3, M=0.5, \phi_{l}^i=-\phi_{l}^f = 0.5, \phi_u^i=-0.9, \phi_u^f=0.6, \gamma_c = 1$. One can verify that this choice of parameters leads to $C_l^i=-C_u^i=1$ and $C_l^f=-C_u^f=-1$. The desired exchange of the Chern numbers between the layers is achieved when the following two conditions are satisfied: (i) the layers are rendered to pure states at the end of the process, i.e., $\Theta_{l(u)}(\tau) = \Tr\left[\rho_{l(u)}(\tau)^2\right]=1$, and (ii) $\tilde{C}_{l(u)}(\tau)=C_{l(u)}^f$. As can be seen from Fig.~\ref{fig_dyn}, which shows the temporal evolution of $\tilde{C}_{l(u)}$ and $\Theta_{l(u)}$, the above conditions are indeed satisfied; the Chern numbers of the layers are therefore exchanged within a finite time $\tau$.\\

Finally, we would like to emphasize here that the protocol presented above does not require the two layers to be conjugate pairs of each other, unlike the protocol presented in Ref.~\onlinecite{barbarino_20}. In the example above, conjugacy would have been maintained if $\phi_l(t)=-\phi_u(t)$ for all $t$, which is clearly not the case. In fact, any desired transformation from a given configuration of Chern numbers to a targeted one can be achieved as long as the total Chern number remains the same throughout the process. Thus, a complete knowledge of the topological phases of the bilayer system allows one to identify viable adiabatic paths to tune the Chern number of the monolayers.\\

{\section{Phase diagram for non-staggered interaction}

As discussed in Sec.~\ref{sec_model}, the advantage of choosing a staggered interaction between the layers is that the band gap can vanish only at the Dirac points, which permits a simpler analysis. However, as shown in Fig.~\ref{fig_gen_phase}, the topological structure of the bilayer Haldane system can appear for other forms of interactions as well. Fig.~\ref{fig_norm_phase} shows the phase diagram for an interaction of the form $\Gamma=\gamma\mathbb{I}$, where $\mathbb{I}$ is the $2 \times 2$ identity matrix. One can identify qualitatively similar topological phases with $C_{tot}=0,\pm1,\pm2$, as those found in the case of a staggered interaction (compare with Fig.~\ref{fig_phase}). The same also holds true for an interaction of the form $\Gamma=\gamma\tau_x$, as shown in Fig.~\ref{fig_inv_phase}. Thus, the topological structure of the bilayer Haldane model is a general feature, irrespective of the exact form of the interaction between the layers. For reference, we have also presented the phase diagram in the case of decoupled layers with $\gamma=0$ in Fig.~\ref{fig_gamma0}.}

{\section{Summary and outlook}\label{sec_summary}

Summarizing, we have shown that a bilayer composite of two coupled Haldane systems possess a robust topological structure exhibiting bulk-boundary correspondences. The topologically protected edge states can either be confined to individual layers or diffused across the edges of the layers, depending on the exact value of the bulk Chern number.} Further, we have also explicitly demonstrated that by dynamically breaking and eventually restoring the ${\rm U(1)}\times{\rm U(1)}$ subgroup of the complete ${\rm U(2)}$ gauge symmetry in a bilayer Chern system, it becomes experimentally viable to realize adiabatic tuning of the Chern phases of the individual layers {even in the thermodynamic limit}; the required protocol for the same is easily identified through a careful inspection of the topological landscape of the bilayer system. {Interestingly, the opening up of a gap in the spectrum due to the coupling between the two layers can be thought to be equivalent to a counter-diabatic process\ct{sels17} suppressing diabatic excitations in the thermodynamic limit. In this regard, optimizing the possible adiabatic pathways to facilitate quicker preparation of non-trivial Chern states can be an interesting problem for future investigations. Moreover, the results presented in this work} are not strictly restricted to a bilayer Haldane system and can be easily generalized to any 2D Chern insulating system.{\scriptsize }\\

{We note in passing that the unitary protocol presented in our work can be analogously compared to the adiabatic tuning of symmetry-protected-topological phases in one-dimensional systems. There, it has been shown\ct{dutta_unitary} that adiabaticity can be maintained throughout, if any of the protecting discrete symmetry (time reversal, charge conjugation or chiral) is broken during the tuning process.} \\

A future direction of study might be to look into the possible topological classifications of twisted bilayer systems and the robustness of the adiabatic protocol discussed in this work in such systems. Further, it may be worthwhile to investigate the topological transitions in more than two connected layers of Haldane-like systems. For example, an immediate generalization of our results can be made for a system with $N$ such layers with finite interlayer couplings, in which case the total Chern number $C_{tot}\in[-N,N]$ of all the layers in the ground state becomes invariant under arbitrary unitary dynamics. {It then follows that for a fixed $N$, there exist $2N+1$ distinct topological sectors of the complete system characterized by the total Chern number, which cannot be adiabatically connected to each other. However, each such sector comprises of several topologically distinct configurations of the individual layers in the decoupled limit. Hence, similar to the bilayer system, it may then be possible to dynamically induce adiabatic transitions (by introducing a finite coupling between the layers) between these multiple topological configurations of the individual layers adding up to the same $C_{tot}$, even in thermodynamically large systems.}
 
\begin{acknowledgments}
We acknowledge HPC-2010, IIT Kanpur, for computational facilities. Sourav Bhattacharjee acknowledges CSIR, India for financial support. Souvik Bandyopadhyay acknowledges financial support from PMRF, MHRD, India. DS acknowledges financial support from DST, India through Project No. SR/S2/JCB-44/2010.
AD acknowledges financial support from a SPARC program, MHRD, India and SERB, DST, New Delhi, India.
\end{acknowledgments}

\appendix

\section{Haldane Model}\label{app_haldane}

\begin{figure}[h]
	\includegraphics[width=0.4\textwidth]{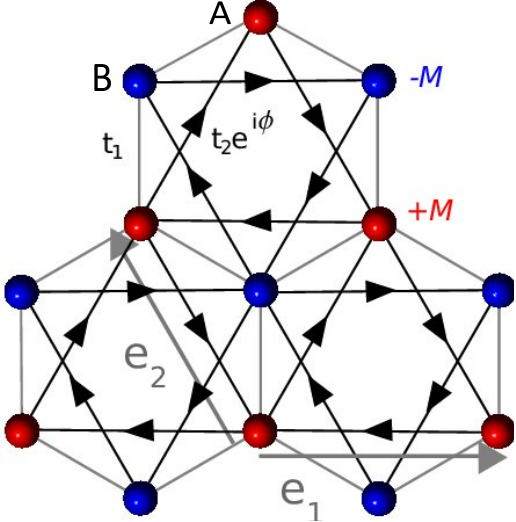}
	\caption{(Figure taken from Ref.~[\ct{haldane_fig}]) Schematic representation of the nearest-neighbor and next-nearest-neighbor couplings in the monolayer Haldane model}\label{fig_hal}
\end{figure}

The Haldane model \ct{haldane83} is an integrable two-dimensional model of spinless electrons. It is based on the graphene honeycomb lattice (Fig.~\ref{fig_strip} of main text) with broken sublattice symmetry (SLS) and time-reversal symmetry (TRS). The Hamiltonian is given by,
\begin{widetext}
\begin{equation}
H=t_1\sum_{i,j=NN}c_{iA}^\dagger c_{jB}^{ } +t_2\sum_{i,j=NNN}e^{i\phi_{ij}}\left(c_{iA}^{\dagger}c_{jA}^{} + c_{iB}^\dagger c_{jB}^{}\right)+M\sum_i\left(c_{iA}^{\dagger}c_{iA}^{} - c_{iB}^\dagger c_{iB}^{}\right) + H.c.,
\end{equation}
\end{widetext}
where $A$ and $B$ identify the two sublattices of the honeycomb lattice, and 
$t_1$ and $t_2$ are the amplitudes of the nearest-neighbor (NN) and 
next-nearest-neighbor (NNN) hoppings, respectively (see Fig.~\ref{fig_hal}). The (TRS) is broken by the complex NNN hoppings, the arguments of which, $\phi_{ij}=\pm \phi$, 
is chosen to be positive (negative) for hoppings in the clockwise (anticlockwise) sense. The SLS, on the other hand, is broken both by the complex hoppings and the Semenoff mass $M$.
Within the bulk, we can assume periodic boundary conditions. The 
Hamiltonian then decouples for each lattice momentum mode within the Brillouin zone (BZ), $H=\bigoplus_{\bf k} {\bm c}_{\bf k}^\dagger H(\bm k){\bm c}^{}_{\bf k}$, where ${\bm c}_{\bf k} = \left(c_{{\bf k},\mathrm{A}}, c_{{\bf k},\mathrm{B}}\right)$. The single-particle Hamiltonian $H(\bm k)$ assumes the Bloch form,
\begin{equation}
H(\bm k)=\bm d(\bm k)\cdot{\bm \sigma}+d_0(\bm k)\mathbb{\textit{I}},
\label{eq_hk}
\end{equation}
where $\bm \sigma\equiv\left(\sigma_x,\sigma_y,\sigma_z\right)$ are the Pauli matrices, $\mathbb{\textit{I}}$ is the $2\times2$ identity matrix, and
\begin{subequations}
	\begin{equation}
	d_x(\bm k)=t_1\left(\cos{(\bm k\cdot{\bm e_1})}+\cos{(\bm k\cdot{\bm e_2})}+\cos{(\bm k\cdot{\bm e_3})}\right),
	\label{eq_hx}
	\end{equation}
	\begin{equation}
	d_y(\bm k)=t_1\left(\sin{(\bm k\cdot{\bm e_1})}+\sin{(\bm k\cdot{\bm e_2})}+\sin{(\bm k\cdot{\bm e_3})}\right),
	\label{eq_hy}
	\end{equation}
	\begin{equation}
	d_z(\bm k)=M-2t_2\sin{\phi}\Big(\sin{(\bm k\cdot{\bm v_1})}+\sin{(\bm k\cdot{\bm v_2})}+\sin{(\bm k\cdot{\bm v_3})}\Big),
	\label{eq_hz}
	\end{equation}
	\begin{equation}
	d_0(\bm k)=-2t_2\cos{\phi}\Big(\cos{(\bm k\cdot{\bm v_1})}+\cos{(\bm k\cdot{\bm v_2})}+\cos{(\bm k\cdot{\bm v_3})}\Big).
	\end{equation}
\end{subequations}
Here, for a given lattice site, the vectors $\{{\bm e_i}\}$ and $\{{\bm v_i}\}$ ($i=1,2,3$) are the locations of the NN and NNN sites respectively. The component $d_0(\bm k)$ has been ignored in the bilayer Haldane model discussed in the main text as it only renormalizes the energy levels of each lattice momentum mode and does not affect the topological properties of the system. The energy spectrum is thus given by
\begin{equation}
E = \pm\sqrt{d_x(\bm k)^2+d_y(\bm k)^2+d_z(\bm k)^2}.
\end{equation}

The topological phases of the Haldane model are characterized by a topological order parameter, namely the Chern number $C$, which takes on only integer quantized values. When $C=0$, the system exists in a trivial phase and behaves as a normal insulator, while for $C=\pm1$, chiral edge states arise which are topologically protected and hence robust, while the bulk of the system remains insulating. The phases are separated from one another by the critical lines, at which the band gap vanishes. Note that when $M=t_2=0, t=1$, the Hamiltonian reduces to that of the gapless graphene Hamiltonian with no topological properties. 

\begin{figure*}
	\subfigure[]{
		\includegraphics[width=0.5\textwidth]{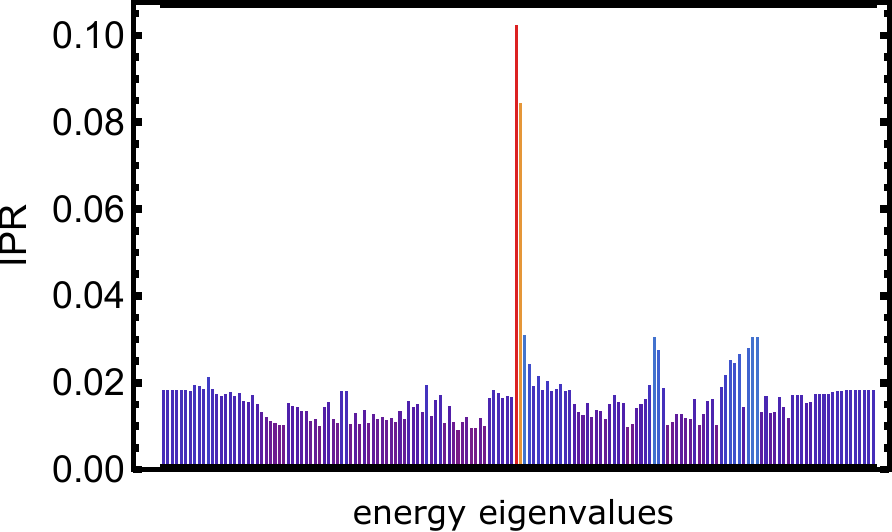}
		\label{fig_ipr1}}%
	\subfigure[]{
		\includegraphics[width=0.5\textwidth]{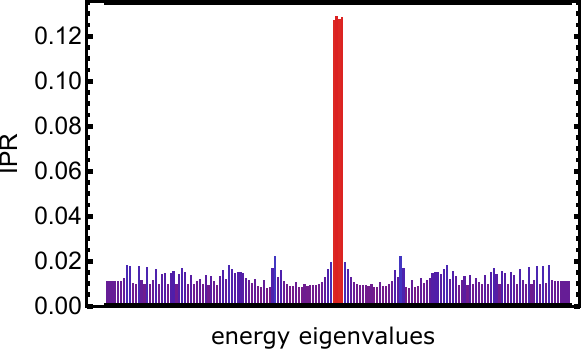}
		\label{fig_ipr2}}	
	\caption{Inverse participation ratio values for all energy eigenstates in a semi-finite bilayer Haldane model for (a)~$C_{tot}=1$ and (b)~ $C_{tot}=2$. The IPR values identify two (one occupied) and four (two occupied) localized edge states in case (a) and (b), respectively.}\label{ipr}
\end{figure*}

\section{Chern number of multiple occupied band systems}\label{app_chern} 

In general, for an $N$-band system with completely filled $N/2$ bands (half-filling), the Berry connection is given by a $N/2\times N/2$ matrix of the form,
\begin{equation}
\bm A_{nm}=i\bra{n(\bm k)}\nabla_{\bf k}\ket{m(\bm k)},
\end{equation} 
where $n,m=1,2,\dots,N/2$, label the energy eigenstates corresponding to the filled energy bands. The tensor form arises due to the fact that multiple filled bands become indistinguishable at degenerate points and the individual Chern numbers fail to remain quantized. Unlike a single filled band, the generic gauge invariance of the quantum state is no longer ${\rm U}(1)$. In the case of the bilayer Haldane system, where $N=4$, a generic gauge transformation in the larger filled subspace (of two bands) takes the form
\begin{equation}
\Psi_{\bf k}^{\prime}=\Psi_{\bf k}\mathcal{U}_{\bf k},
\end{equation}
where $\Psi_{\bf k}=\left(\ket{\phi^1_{\bf k}}~\ket{\phi^2_{\bf k}}\right)$ is a spinor comprising of the two occupied states $\ket{\phi^1_{\bf k}}$ and $\ket{\phi^2_{\bf k}}$, and $\mathcal{U}_{\bf k}$ is an arbitrary ${\rm U}(2)$ matrix. Importantly, the gauge group ${\rm U}(2)$ is not Abelian, and hence, one needs 
to define a non-Abelian connection for parallel transport of the spinors in 
this space.

The non-Abelian curvature form $F_{na}$ is given by,
\begin{equation}
F_{na}^{ij}(\bm k)=\frac{\partial}{\partial k_i}A^j-\frac{\partial}{\partial k_j}A^i-i\left[A^i,A^j\right],
\end{equation}
where $i,j$ denote the components of the vectors along the unit vectors of the reciprocal lattice. The Chern number characterizing the topological phase of the system is then calculated as follows \ct{wilczek84},
\begin{equation}
C=\frac{1}{2\pi}\int_{BZ}d^2k~\Tr \left(F_{na}(\bm k)\right).
\end{equation}
A convenient form of the Chern number can be derived from the above equation in terms of the projection operator $P$ on the ground state manifold of the completely filled bands \ct{murakami04},
\begin{equation}\label{app_eq_chern}
C = \frac{i}{2\pi}\int_{BZ}d^2k~\Tr\left(P\left[\frac{\partial P}{\partial k_i}, \frac{\partial P}{\partial k_j} \right]\right), 
\end{equation}
where $P = \sum_{n=1}^{N/2}\ket{n(\bm k)}\bra{n(\bm k)}$. For $N=2$, we recover the commonly used definition of the Chern number, which characterizes the topological phases of two-band Chern insulators. Further, if all the occupied bands are gapped among themselves, the $C$ is equivalent to the sum of the Chern numbers of the occupied bands. On the contrary, if any degeneracy arises among the occupied bands, the Chern numbers of the individual bands no longer remain well-defined. The total Chern number, as defined in Eq.~\eqref{app_eq_chern} however, remains integer quantized as long as the gap between the occupied and the empty bands remains finite. 

\section{Inverse participation ratios of the energy eigenstates for semi-open boundary conditions}\label{app_ipr}

We consider the semi-infinite bilayer Haldane system with armchair edges discussed in the main text. The inverse participation ratio (IPR) of an energy eigenstate $\psi_n$ is defined as,
\begin{equation}
\mathrm{IPR}(\psi_n) = \sum_{m_l,s}\left|\langle m_l^s|\psi_n\rangle\right|^4+\sum_{m_u,s}\left|\langle m_u^s|\psi_n\rangle\right|^4,
\end{equation}
where $m_{l(u)}$ is the `strip' index for the lower (upper) layer and $s$ is the sublattice index. If a given eigen-state is extended in real space, then one can roughly assume $\left|\langle m_{l(u)}^s|\psi_n\rangle\right|^2\approx1/2M$, where $M$ is the total number of horizontal strips. The IPR for extended states thus diminishes with increasing $M$ and vanish in the thermodynamic limit. On the other hand, for localized states, the IPR remains finite with increasing $M$.

In Fig.~\ref{fig_ipr1}, we plot the IPR of all energy eigenstates when $C_{tot}=1$ with $M=40$. It is clearly seen that the IPR is significantly higher for a pair of eigen-states, confirming that there exist a pair of localized edge states, only one of which is occupied in the ground state of the system. Similarly, Fig.~\ref{fig_ipr2} shows the presence of four localized states when $C_{tot}=2$, two of which are occupied in the ground state. Hence, there exists a one-to-one correspondence between the number of occupied edge states and the total Chern number.

\section{Chern number from two-particle ground state}\label{app_2p}

As the two negative energy bands are completely filled at half-filling and 
the total particle number is conserved, the ground state of the bilayer system resides in the two-particle Hilbert space. Within this restricted Hilbert space, two out of the four single-particle states are occupied for each lattice momentum $\bm k$. The basis states can thus be constructed as $\{c_{{\bf k},\mathrm{A}}^{l\dagger}c_{{\bf k},\mathrm{B}}^{l\dagger}\ket{0}$, $c_{{\bf k},\mathrm{A}}^{l\dagger}c_{{\bf k},\mathrm{A}}^{u\dagger}\ket{0}$, $c_{{\bf k},\mathrm{A}}^{l\dagger}c_{{\bf k},\mathrm{B}}^{u\dagger}\ket{0}$, $c_{{\bf k},\mathrm{B}}^{l\dagger}c_{{\bf k},\mathrm{A}}^{u\dagger}\ket{0}$, $c_{{\bf k},\mathrm{B}}^{l\dagger}c_{{\bf k},\mathrm{B}}^{u\dagger}\ket{0}$, $c_{{\bf k},\mathrm{A}}^{u\dagger}c_{{\bf k},\mathrm{B}}^{u\dagger}\ket{0}\}$, where $\ket{0}$ represents the zero-particle vacuum state. The two-particle Hamiltonian in this basis is given by
\begin{widetext}
\begin{equation}
H(\bm k)=\begin{pmatrix}
0 & 0 & -\gamma & -\gamma & 0 & 0\\
0 & d_z^l(\bm k)+d_{z}^u(\bm k) & d_x(\bm k)-id_y(\bm k) & d_x(\bm k)-id_y(\bm k) & 0 & 0\\
-\gamma & d_x(\bm k)+id_y(\bm k) & d_{z}^l(\bm k)-d_{z}^u(\bm k) & 0 & d_x(\bm k)-id_y(\bm k) & \gamma\\
-\gamma & d_x(\bm k)+id_y(\bm k) & 0 & -d_{z}^l(\bm k)+d_{z}^u(\bm k) & d_x(\bm k)-id_y(\bm k) & \gamma\\
0 & 0 & d_x(\bm k)+id_y(\bm k) & d_x(\bm k)+id_y(\bm k) & -d_{z}^l(\bm k)-d_{z}^u(\bm k) & 0\\
0 & 0 & \gamma & \gamma & 0 & 0
\end{pmatrix}.
\end{equation}
\end{widetext}

The two-particle energy bands can be obtained by diagonalizing the above 
Hamiltonian. The Chern number can then be calculated by integrating the 
(Abelian) Berry curvature of the lowest energy band over the BZ. The two-particle Chern number $C_{2p}$ thus calculated is in fact equivalent to the total Chern number described in the main text. To see this explicitly, we write the two-particle ground state as $\ket{\psi} = \otimes\ket{\psi(\bm k)}=\otimes\ket{\phi_1(\bm k)}\ket{\phi_2(\bm k)}$ , where $\ket{\phi_1(\bm k)}$ and $\ket{\phi_2(\bm k)}$ are the negative energy single-particle states. The Berry connection is found to be
\begin{align}
\bm A(\bm k)&=i\bra{\psi(\bm k)}\nabla_{\bf k}\ket{\psi(\bm k)}\non\\&=i\bra{\phi_1(\bm k)}\nabla_{\bf k}\ket{\phi_1(\bm k)}+i\bra{\phi_2(\bm k)}\nabla_{\bf k}\ket{\phi_2(\bm k)}\non\\&={\bm A}_{11}(\bm k)+{\bm A}_{22}(\bm k),
\end{align}
and the Berry curvature is obtained as
\begin{equation}\label{eq_ab_berry}
F(\bm k) = \frac{\partial}{\partial k_i}A_{11}^j-\frac{\partial}{\partial k_j}A_{11}^i+\frac{\partial}{\partial k_i}A_{22}^j-\frac{\partial}{\partial k_j}A_{22}^i.
\end{equation}
The two-particle Chern number is then calculated as
\begin{equation}
C_{2p}=\frac{1}{2\pi}\int_{BZ}d^2k~F(\bm k).
\end{equation}
On the other hand, the non-Abelian Berry curvature is given by (see Eq.~(6) of the main text),
\begin{equation}
F_{na}(\bm k)=\frac{\partial}{\partial k_i}A^j-\frac{\partial}{\partial k_j}A^i-i\left[A^i,A^j\right],
\end{equation} 
It is straightforward to check that $\Tr[A^i, A^j]=0$ and thus,
\begin{equation}
\Tr \left(F_{na}(\bm k)\right) = \frac{\partial}{\partial k_i}A_{11}^j-\frac{\partial}{\partial k_j}A_{11}^i+\frac{\partial}{\partial k_i}A_{22}^j-\frac{\partial}{\partial k_j}A_{22}^i = F(\bm k),
\end{equation}
where $F(\bm k)$ is the (Abelian) Berry curvature in Eq.~\eqref{eq_ab_berry}. Hence, the Chern number is found to be
\begin{equation}
C_{tot} = \frac{1}{2\pi}\int_{BZ}d^2k~\Tr \left(F_{na}(\bm k)\right) = C_{2p}.
\end{equation}

\end{document}